\documentclass[fleqn,usenatbib]{mnras}

\usepackage{newtxtext,newtxmath}
\usepackage[T1]{fontenc}
\usepackage{ae,aecompl}

\usepackage{graphicx}	
\usepackage{amsmath}
\usepackage{amssymb}

\title[$X_{\rm CO}$ in cosmological simulations]{Reproducing the CO-to-H$_2$ conversion factor in cosmological simulations of Milky Way-mass galaxies}

\author[L.C. Keating et al.]{\parbox{\textwidth}{Laura C. Keating$^{1}$\thanks{E-mail: lkeating@cita.utoronto.ca}, Alexander J. Richings$^{2}$, Norman Murray$^{1,3}$,\\ Claude-Andr\'e Faucher-Gigu\`ere$^{4,5}$, Philip F. Hopkins$^6$, Andrew Wetzel$^{7}$,\\ Du\v san Kere\v s$^8$, Samantha Benincasa$^{7}$, Robert Feldmann$^{9}$, Sarah Loebman$^7$\thanks{Hubble Fellow}\\ and Matthew E. Orr$^6$}\vspace{0.4cm} \\
\parbox{\textwidth}{$^1$Canadian Institute for Theoretical Astrophysics, 60 St. George Street, University of Toronto, ON M5S 3H8, Canada\\
$^2$ Institute for Computational Cosmology, Department of Physics, Durham University, South Road, Durham, DH1 3LE, United Kingdom\\
$^3$ Canadian Research Chair in Theoretical Astrophysics\\
$^4$ Center for Interdisciplinary Exploration and Research in Astrophysics (CIERA)\\
$^5$ Department of Physics and Astronomy, Northwestern University, 2145 Sheridan Road, Evanston, IL 60208, USA\\
$^6$ TAPIR, MC 350-17, California Institute of Technology, Pasadena, CA 91125, USA\\
$^7$ Department of Physics, University of California, Davis, CA 95616, USA\\
$^8$ Department of Physics, Center for Astrophysics and Space Sciences, University of California at San Diego, La Jolla, CA 92093\\
$^9$ Institute for Computational Science, University of Zurich, Zurich CH-8057, Switzerland\\
}
}

\begin{document}

\date{\today}

\pagerange{\pageref{firstpage}--\pageref{lastpage}} 

\pubyear{2019}

\maketitle

\label{firstpage}

\begin{abstract}
We present models of CO(1-0) emission from Milky Way-mass galaxies at redshift zero in the FIRE-2 cosmological zoom-in simulations. We calculate the molecular abundances by post-processing the simulations with an equilibrium chemistry solver while accounting for the effects of local sources, and determine the emergent CO(1-0) emission using a line radiative transfer code. We find that the results depend strongly on the shielding length assumed, which in our models sets the attenuation of the incident UV radiation field. At the resolution of these simulations, commonly used choices for the shielding length, such as the Jeans length, result in CO abundances that are too high at a given H$_2$ abundance. We find that a model with a distribution of shielding lengths, which has a median shielding length of $\sim 3$ pc in cold gas ($T < 300$ K) for both CO and H$_{2}$, is able to reproduce both the observed CO(1-0) luminosity and inferred CO-to-H$_{2}$ conversion factor at a given star formation rate compared with observations. We suggest that this short shielding length can be thought of as a subgrid model which controls the amount of radiation that penetrates giant molecular clouds.
\end{abstract}


\section{Introduction}

Understanding the molecular gas content of galaxies is crucial for understanding the process of star formation \citep[e.g.,][]{mckee2007,kennicutt2012,krumholz2014}. Molecular gas tends to form in high density regions, where self-shielding and shielding by dust from the interstellar radiation field is most effective. Much of this molecular gas is in the form of giant molecular clouds (GMCs), which typically have average densities larger than 10$^{2}$ cm$^{-3}$ and temperatures in the range 10--20 K \citep{ferriere2001}. The most abundant molecule is molecular hydrogen (H$_{2}$). However, it is challenging to observe the H$_{2}$ in emission, as it requires high temperatures ($T \sim 500$ K) to be excited and therefore cannot be seen in emission at the low temperatures of GMCs. Instead, carbon monoxide (CO) can be used as a convenient tracer of this molecular gas, as it has its first rotational transition at 5.5 K \citep{carilli2013} - low enough to probe the cold interstellar medium (ISM). 

However, it has long been known that CO is not a perfect tracer of H$_{2}$, as CO is more easily dissociated than H$_{2}$ \citep{vandishoeck1988,wolfire2010}. This results in regions of ``dark gas'' in a phase that is lower in density and warmer than the CO-bright gas \citep[see, e.g., the model of][]{seifreid2019}. Dark gas can be probed observationally by $\gamma$-rays produced from interactions between cosmic rays and hydrogen \citep{grenier2005,remy2017}, with maps of thermal dust emission \citep{planck2011darkgas} and from emission from singly ionized carbon atoms \citep{pineda2013,langer2014}. 

The CO(1-0) emission is usually related to the total H$_{2}$ mass by a CO-to-H$_{2}$ conversion factor $X_{\rm CO}$, defined as
\begin{equation}
\label{eqn:xco}
  X_{\rm CO} = \frac{N_{\rm H_{2}}}{W_{10}},
\end{equation}
where $N_{\rm H_{2}}$ is the H$_{2}$ column density and $W_{10}$ is the velocity-integrated CO(1-0) brightness temperature, which is related to the CO(1-0) intensity. In the disc of the Milky Way, this conversion factor is estimated to be $ X_{\rm CO} \approx 2 \times 10^{20}$ cm$^{-2}$ K$^{-1}$ km$^{-1}$ s using a range of different techniques \citep{bolatto2013}. This conversion factor is however also thought to depend on environmental factors, such as the metallicity \citep{israel1997,leroy2011,sandstrom2013}. The most appropriate choice for a given galaxy, or whether a single value for an individual galaxy is valid, is a subject of much observational \citep{solomon1987,downes1998,accurso2017} and theoretical discussion \citep{wolfire1993,narayanan2011,narayanan2012,shetty2011a,shetty2011b,feldmann2012,feldmann2012b,gong2018,li2018,richings2018}.

One way of understanding how CO emission relates to properties of the ISM and galaxy is through simulations. However, modelling the emission of CO from galaxies is extremely challenging, due to the wide range of spatial scales involved. CO has been observed in the Milky Way tens of kpc from the galactic centre \citep{heyer2001}, but the molecular gas resides in GMCs with sizes of a few to $\sim$100 pc \citep{mivilledeschenes2017}. These clouds further contain clumps and cores that have sizes less than 0.1 pc \citep{bergin2007}. Some works have therefore focused on modelling individual GMCs \citep[e.g.,][]{glover2011,shetty2011a,shetty2011b,clark2015,penaloza2018}. This allows for high spatial resolution in the regions hosting the molecular gas, but this comes at the cost of having to neglect the impact of the larger scale galactic environment. Another approach is to model the emission coming from entire or parts of isolated galaxies \citep[e.g.,][]{duartecabral2015,glover2016,richings2016,gong2018}. This still neglects the fact that galaxy formation occurs in a cosmological context. At the largest scales, semi-analytic schemes can be used to quickly compute CO luminosities for large numbers of galaxies \citep{obreschkow2009, lagos2012, popping2019}, which provide large samples of CO emitting galaxies but cannot be used to study small scale environmental effects. An intermediate approach is to post-process cosmological hydrodynamic simulations by making some assumptions about the internal structure of GMCs \citep{olsen2016,vallini2018,li2018}. Ideally one would calculate the molecular gas abundances on the fly. However, such simulations are extremely costly due to the increased number of species that need to be tracked and rate equations that must be integrated. These simulations have so far been limited to high-redshift studies \citep{katz2017,pallottini2019,lupi2019b}, isolated galaxies \citep{kannan2019} or dwarf galaxies \citep{lupi2019}.

In this paper we continue the attempt to bridge the gap between simulations of individual GMCs or isolated galaxies, and fully cosmological simulations. We use cosmological zoom-in simulations from the FIRE project\footnote{https://fire.northwestern.edu/} \citep[Feedback in Realistic Environments;][]{hopkins2014fire,hopkins2018fire2}. These simulations have high mass resolution and an explicit treatment of the multi-phase ISM, and can resolve the most massive GMCs ($M \sim 10^5 \, M_{\odot}$, \citealt{benincasa2019}) where the bulk of star formation occurs \citep{williams1997}. We post-process the redshift zero simulation outputs with an equilibrium chemistry solver and a line radiative transfer code to calculate the CO and  H$_{2}$ abundances, and the CO(1-0) emission. The structure of this paper is as follows. In section 2 we describe how we model the CO(1-0) line emission in our simulations. In section 3 we compare our models with observations of local galaxies and describe the effects of varying some of the assumptions in our chemical modelling. In section 4 we discuss the interpretation of our results. We present our conclusions in section 5.

\section{Modelling CO(1-0) emission}

\subsection{Cosmological zoom simulations of Milky Way-type galaxies}

In this work, we make use of the FIRE-2 simulations first presented in \citet{hopkins2018fire2}. The FIRE-2 simulations are a set of cosmological zoom-in simulations that explicitly model a multi-phase ISM. The simulations were run with \textsc{gizmo}, using the meshless-finite mass hydrodynamic solver \citep{hopkins2015gizmo}. \textsc{gizmo} uses a gravity solver and domain decomposition algorithm descended from the Tree-PM solver in \textsc{gadget-3} \citep[last described in][]{springel2005gadget2} with some modifications, including adaptive softening for gas.

The physics included in the FIRE-2 simulations is described in detail in \citet{hopkins2018fire2} and only a brief summary is provided here. Star formation occurs using a probabilistic criteria for gas that is self-gravitating, Jeans-unstable, molecular and has density $n > 1000$ cm$^{-3}$. During the course of the simulation, the molecular fraction of the gas particles is determined using the relation presented in \citet{krumholz2011}. Each star particle is treated as a single stellar population assuming a \citet{kroupa2001} initial mass function. The stellar feedback of each star particle is estimated from tables computed using the stellar population models of \textsc{starburst99} \citep{leitherer1999}. The stellar feedback mechanisms include Type-Ia and Type-II supernovae, stellar winds from evolved stars, photoionization/-heating and radiation pressure. Photoionization and photoheating from the extragalactic UV background is also included \citep{fauchergiguere2009}, as well as photoelectric heating and heating by cosmic rays. Heating and cooling rates are estimated from pre-computed \textsc{cloudy} tables \citep{ferland2013} that include metal-line and molecular cooling, allowing the gas to cool down to 10 K. Eleven elements are explicitly tracked (H, He, C, N, O, Ne, Mg, Si, S, Ca and Fe). Subgrid metal mixing is included using the prescription described in \citet{hopkins2017md} and \citet{escala2018}.

\begin{figure*}
  \includegraphics[width=2\columnwidth]{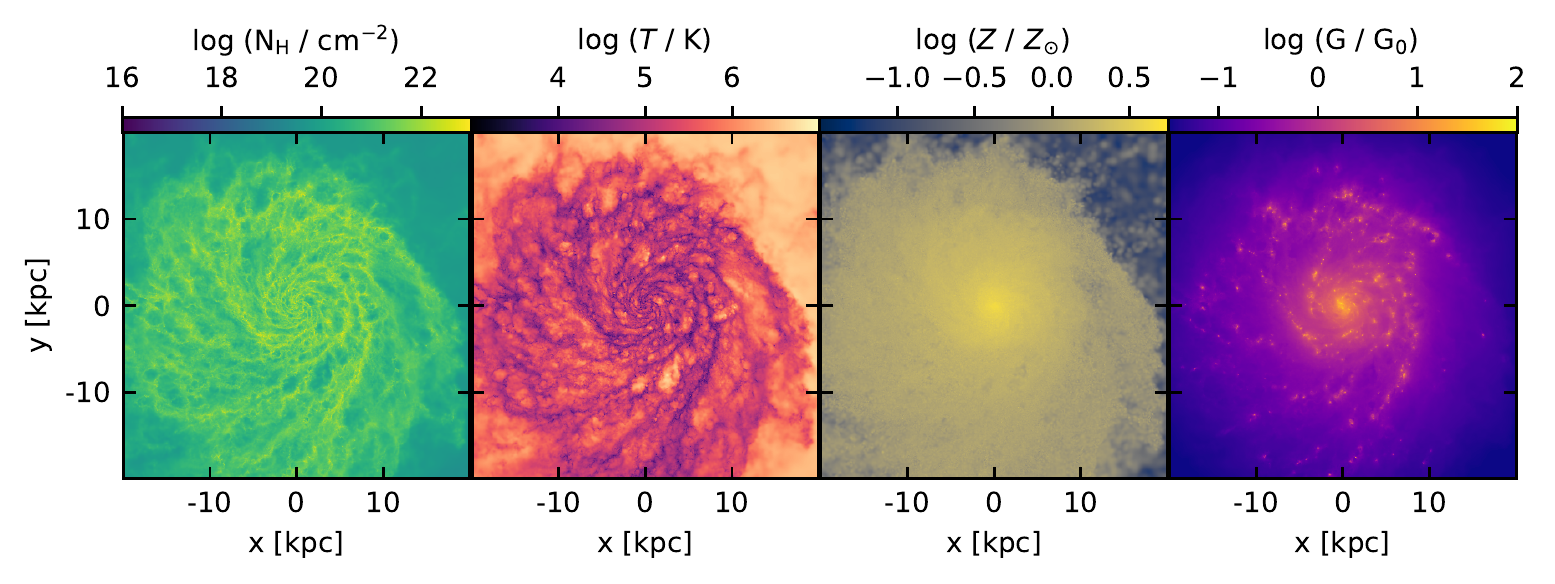}
  \caption{First row, from left to right: Maps of the total hydrogen density, the mass-weighted gas temperature, the mass-weighted metallicity in solar units and mass-weighted UV flux in Habing units for the m12i simulation.}
  \label{gal_maps}
  \end{figure*}

We focus most of our analysis here on the m12i galaxy, presented as part of the Latte suite of FIRE-2 simulations in \citet{wetzel2016}. This is a simulation of a Milky Way-mass galaxy, with a virial mass of $1.1 \times 10^{12} \, M_{\odot}$. Maps of the projected density, mass-weighted temperature, mass-weighted metallicity and mass-weighted UV flux we measure in m12i are shown in Figure \ref{gal_maps}. The halo was selected for its final mass from a dark matter only simulation of a 85.5 Mpc volume, with initial conditions generated using \textsc{music} \citep{hahn2011music} as part of the AGORA project \citep{kim2014}. A $\Lambda$CDM cosmology was assumed, with $\Omega_{\Lambda}=0.728$, $\Omega_{\rm m}=0.272$, $\Omega_{\rm b}=0.0455$, $h=0.702$, $\sigma_8=0.807$ and $n_{\rm s}=0.961$. The gas elements have mass 7100 $M_{\odot}$ and the force resolution reaches pc scales in gas resolution elements with densities a few times 10$^{3}$ cm$^{-3}$. Using the FIRE-2 physics model described above, this simulation has been shown to agree with many observed properties of our Galaxy, such as the properties of satellite galaxies \citep{wetzel2016}, the stellar mass-halo mass relation \citep{hopkins2018fire2}, the mass-metallicity relation \citep{ma2017} and the Kennicutt-Schmidt relation \citep{orr2018} and properties of the stellar thin and thick disc \citep{sanderson2018}, among others. Most relevant for this work, the simulation also reproduces many observed properties of GMCs (\citealt{guszejnov2019}, \citealt{benincasa2019} and Lakhlani et al., in prep.).

\subsection{Chemical equilibrium modelling}
\label{chimes}

As we wish to compute the H$_2$ and CO abundances, we post-process the simulation snapshots with the chemistry solver \textsc{chimes} \citep{richings2014thin,richings2014shielded}. \textsc{chimes} is capable of computing non-equilibrium abundances, but in this work we use it in equilibrium mode (i.e., we evolve the network until chemical equilibrium is reached). One benefit of using \textsc{chimes} is that the same library can be used to calculate equilibrium chemical abundances by post-processing simulation outputs, as well as calculating the full non-equilibrium abundances on the fly in simulations. This allows for a direct comparison between equilibrium and non-equilibrium methods \citep[e.g.,][]{richings2016}. However, calculating the abundances on the fly adds a substantial computational cost and so we only present results assuming equilibrium here. \textsc{chimes} calculates the chemical abundances of 157 species. These include the ionization states of the 11 elements tracked in the simulation, as well as 20 molecular species. The \textsc{chimes} chemical network includes collisional reactions, photochemical reactions, ionization by cosmic rays and dust grain reactions.  We do not evolve the temperature of the gas particles, since the FIRE-2 simulations already account for molecular cooling. This potentially introduces an inconsistency into our analysis, as the metal line and molecular cooling tables used in the hydrodynamic simulation assume that the gas is optically thin but in our chemical modelling the gas is allowed to self-shield. However, we confirmed that also allowing the temperature to evolve to equilibrium along with the chemical abundances did not change our results significantly, with the mass of CO differing by a factor of two between the two cases. As we will discuss later, this is a much smaller change than we see when changing other parameters in our modelling (in particular the shielding length), which can vary the results by orders of magnitude.

One assumption that goes into our chemical modelling is the choice of UV background, which regulates the photoionization and photodissociation rates in the gas. We compute the UV background due to local sources (which dominates over the extragalactic UV background in the disc of the galaxy) by restarting each snapshot of the underlying hydrodynamic simulation and printing out the far-UV fluxes. These are computed using the LEBRON radiative transfer approximation included in the FIRE-2 simulations \citep{hopkins2018fire2,hopkins2019lebron}. The emissivity of the sources is calculated by computing the luminosity of each star particle based on its age, metallicity and mass using the \textsc{starburst99} stellar population models. The far-UV fluxes are defined as the $6-13.6$ eV band, so this is not the ionizing radiation field. However, it gives us an estimate of the inhomogeneous UV background we are interested in. Attenuation in the vicinity of the source by dust is accounted for in the simulation, and the radiation is then transported under the assumption that the gas is optically thin. When the photoionization rates are calculated in the simulation, the radiation is attenuated again for each gas particle to account for local self-shielding. We print out the fluxes before this step, as this is already accounted for in \textsc{chimes} (see below).

Using these UV fluxes, we can now account for spatial variations in the amplitude of the UV background due to local sources. We do not however account for spatial fluctuations in the shape of the background, and use photoionization cross-sections calculated using the shape of the \citet{black1987} interstellar radiation field everywhere. This means we are not accounting for the effects of spectral hardening, and changes in the shape of the spectrum due to different stellar populations. 

We assume that the cosmic ray ionization rate scales linearly with the spatially varying far-UV flux, and use a fiducial cosmic ray ionization rate $\zeta_{\ion{H}{i}}= 1.8 \times 10^{-16}$  s$^{-1}$ \citep{indriolo2012} at the point where the far-UV flux is equal to the average Milky Way value. We also explore the effect of raising/lowering the  cosmic ray ionization rate, as well as a model where the far-UV flux and cosmic ray ionization rate are held constant (described in Section \ref{sec:params}).

In \textsc{chimes}, each species $i$ is shielded from the UV background by a shielding factor calculated based on the local column density. This shielding regulates the photoionization and photodissociation rate of that species. However, in the simulations only a volume density for each particle is tracked. We therefore need to relate the volume density to a column density. We use the definition
\begin{equation}
  N_{i} = n_{i} L_{\rm shield},
\end{equation}
where $N_{i}$ is the column density of species $i$, $n_{i}$ is the number density of the particle and $L_{\rm shield}$ is the shielding length. We assume that $L_{\rm shield}$ is constant for all species, but note that in reality different species may be associated with different shielding lengths, depending on how they are distributed throughout the galaxy (e.g. the molecular gas may be clumpier than the dust).  We will discuss how $L_{\rm shield}$ is chosen in more detail in the following section. Once this column density is known, the shielding factor for each species is calculated and used to attenuate the optically thin rates. This provides us with the rates that are appropriate for optically thick gas. Shielding by dust is accounted for each species $i$ using the shielding factor 
\begin{equation}
S^i_{\rm dust} = \exp(-\gamma^i_{\rm dust} A_{\rm v}),
\end{equation}
where $\gamma^i_{\rm dust}$ is a constant taken from \citet{vandishoeck2006} and \citet{glover2010}. $A_{\rm v}$ is the extinction, defined as $A_{\rm v} = 4.0 \times 10^{-22} (N_{\rm H} /{\rm cm}^{-2}) (Z / Z_{\odot})$ mag cm$^2$. The species we are most interested in here are H$_{2}$ and CO. H$_{2}$ can additionally self-shield, and this is accounted for in a temperature-dependent relation outlined in detail in  \citet{richings2014shielded}. Additional Doppler broadening with line width 7.1 km s$^{-1}$ due to subgrid turbulence at the scale of the resolution element is assumed \citep{krumholz2012} when calculating the H$_{2}$ shielding factor. For CO, self-shielding is taken into account and there is also a contribution from shielding due to H$_{2}$. These shielding factors are taken from \citet{visser2009}.

\subsection{Line radiative transfer}

Once we have the CO abundances, the final step is to calculate the observed CO luminosity. We focus here on the $J=1-0$ transition at 2.6 mm. We compute the luminosity by further post-processing the simulations using the non-LTE line radiative transfer code \textsc{radmc-3d} \citep{dullemond2012radmc}. We perform the radiative transfer on an AMR grid, refining when more than two particles are present in a cell. We use a cubic spline kernel to interpolate the resolution elements onto the grid. We take a 40 kpc cube, centred on the halo, and perform the radiative transfer after rotating the galaxy to be face-on. We perform the radiative transfer twice for each model: once with the line emission plus the dust continuum, and again with the dust continuum alone. Subtracting these provides us with the CO(1-0) emission only. Our final maps have 2048$^{2}$ cells in the spatial plane, with 19.5 pc resolution, and 1 km s$^{-1}$ resolution along the frequency axis.

To calculate the level populations, we use the large velocity gradient method mode in \textsc{radmc-3d} \citep{castor1970,goldreich1974,shetty2011a}. This calculates an optical depth for each cell based on the gas velocities in neighbouring cells, assuming that a photon will eventually be Doppler shifted away from line centre and hence yielding an escape probability for each photon. Gas can also be excited based on its temperature. We use Einstein and collisional rate coefficients taken from the LAMDA database \citep{schoier2005} and also account for the effects of the CMB.

As dust is not explicitly modelled as a separate species in these simulations, we assume that a fraction of the metals in gas with temperature less than 10$^{5}$ K are in dust and use fixed dust to metal gas mass ratios for the silicate and graphite species. We note that we have not implemented a correction for dust depletion into our modelling, which means that we are double-counting metals that would otherwise not be in the gas phase. Estimates of dust depletion factors \citep{jenkins2009,decia2016} suggest that carbon and oxygen can be depleted by $\sim40-60$ per cent, relevant for the CO abundances we are interested in here. This could potentially reduce our modelled CO(1-0) luminosities by a factor $\sim2$. This is not at the level which will change any of the main conclusions of this paper (which, as described later, seeks to resolve order of magnitude discrepancies in the line emission) and is something that we will investigate more carefully in future work. We account for local turbulent broadening by adding a microturbulent line width of 7.1 km s$^{-1}$, consistent with what is assumed in section \ref{chimes}. We assume an external radiation field while performing the radiative transfer. The role of this background radiation is to calculate the dust temperatures and continuum emission. As we will later subtract away the continuum emission to isolate the line emission, we do not use an inhomogeneous UV background for this step (in contrast with our chemical modelling as described in section \ref{chimes}). We instead use a fixed uniform background with the shape and amplitude of the \citet{black1987} model for the interstellar radiation field.

\section{CO(1-0) Emission and Shielding Length}

\begin{figure}
  \includegraphics[width=\columnwidth]{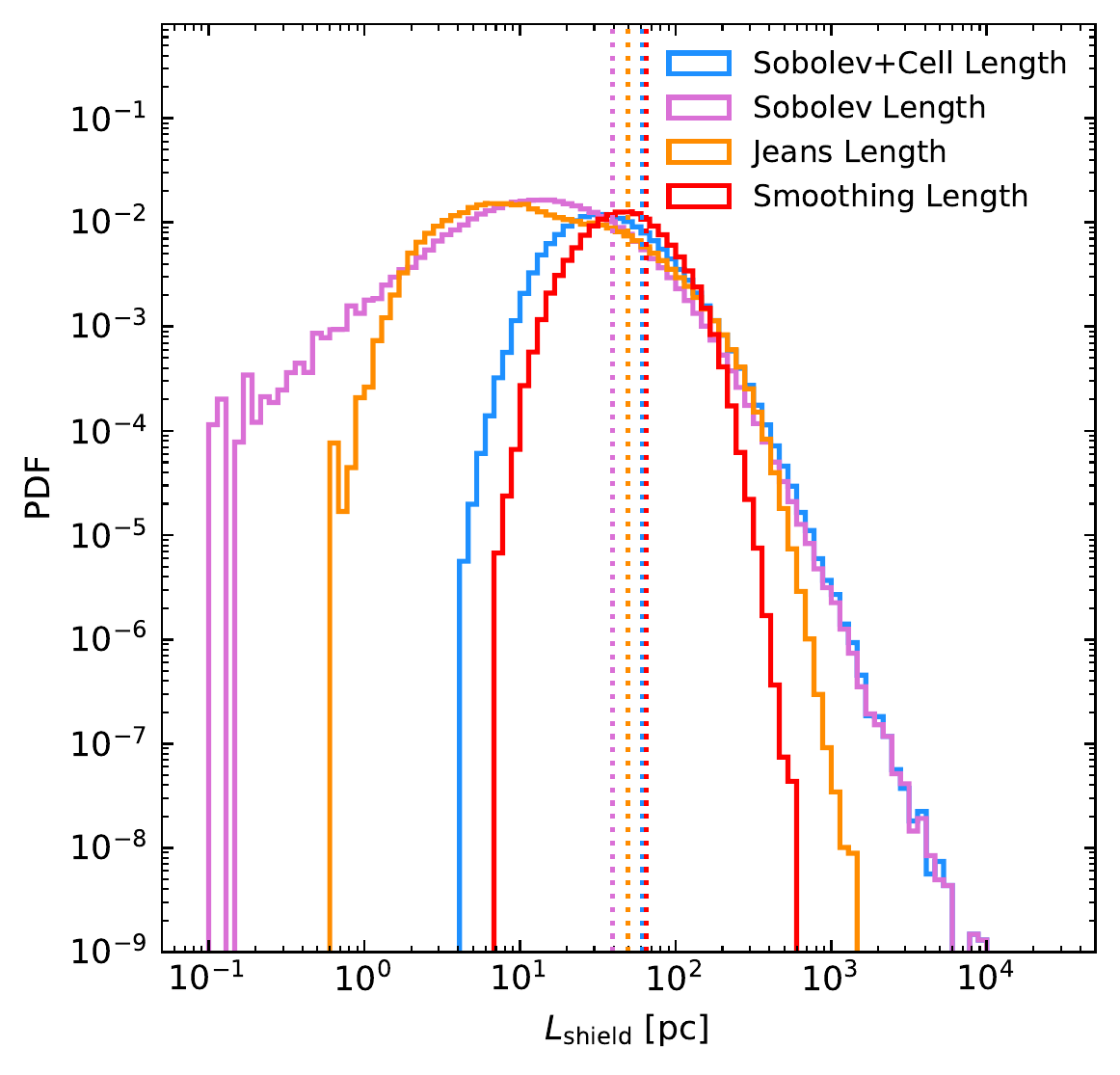}
  \caption{Histograms of the different fiducial shielding length scales we consider here: the smoothing length (red), the Jeans length (orange), the Sobolev length (blue) and the Sobolev length plus a contribution from the local cell (pink). Only values for cold gas ($T < 300$ K) in the galactic disc of m12i (which we define as gas with radius $r < 20$ kpc and vertical height $\lvert z \rvert < 3$ kpc) are shown. The vertical dotted lines mark the median of the distributions.}
  \label{shielding_lengths}
    \end{figure}

\subsection{Shielding length approximations}

As discussed above, one of the fields that must be input to  \textsc{chimes} is the shielding length of each particle ($L_{\rm shield}$), which controls whether the optically thick or optically thin photodissociation rates should be applied. Ideally one would use a full ray-tracing scheme, however this is computationally expensive so an alternative is to assume some local approximation.  However, it is not obvious how this approximate shielding length should be defined. This uncertainty has been noted in previous works, exploring molecular gas formation in different regimes. In a network that includes shielding by dust, H$_2$ and CO self-shielding, and shielding of CO by H$_2$, \citet{safranekshrader2017} performed an analysis of seven different approximations of this shielding length. They compared the resulting H$_2$ and CO abundances to a full ray-tracing simulation. They found that using the Jeans length with a temperature ceiling of 40 K provided the closest solution to the ray-tracing scheme. However, \citet{wolcottgreen2011}, again comparing to full 3D ray-tracing, found that the Sobolev length was a better approximation, although this was in the case of H$_2$ self-shielding alone in gas of primordial composition. There is therefore no consensus on the best approximation to make, and we note that this likely depends on the properties of the simulation, in particular the resolution, as well as the physics that is included. Here we also take the approach of testing different approximations of the shielding length (as well as other parameters assumed in the chemical modelling), and compare the results directly against observations.

\begin{figure*}
  \includegraphics[width=2\columnwidth]{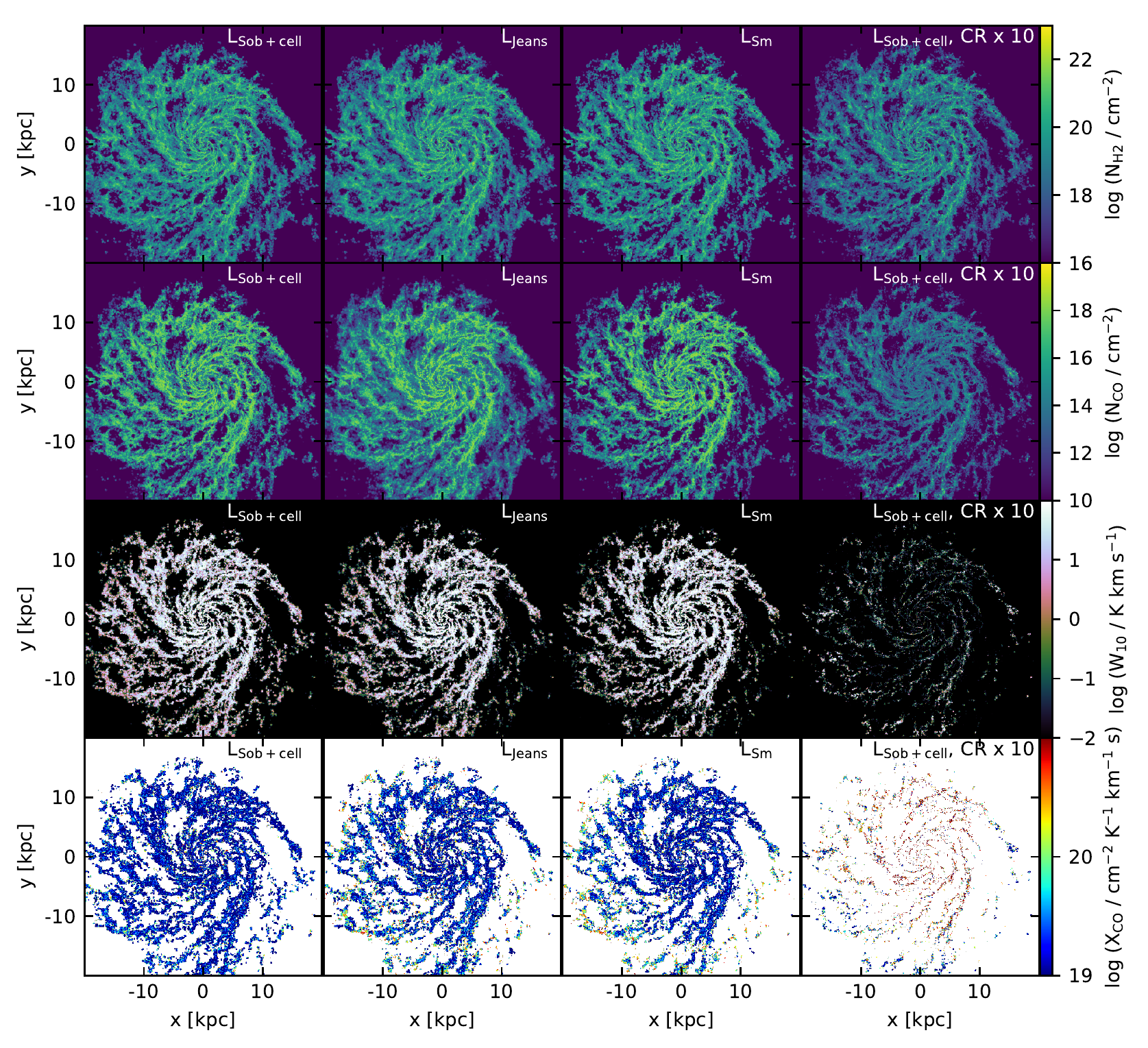}
  \caption{Maps of different quantities resulting from our chemical modelling and line radiative transfer in the m12i simulation, in which we have explored different definitions of the shielding length and normalisations of the cosmic ray ionization rate we assume. First row: The H$_{2}$ column density for four cases: assuming the shielding length is the Sobolev length plus a contribution from the local cell, assuming the shielding length is the Jeans length, assuming the shielding length is the smoothing length and assuming the shielding length is the Sobolev length plus a contribution from the local cell, but also increasing the cosmic ray ionization rate by a factor 10. Second row: The CO column density for the four cases shown in the second row. Third row: The velocity-integrated CO(1-0) brightness temperature for the four cases shown in the second row. Fourth row: The CO-to-H$_{2}$ conversion factor ($X_{\rm CO}$) for the four cases shown in the second row for pixels that have $W_{10} > 0.1$ K km s$^{-1}$.}
  \label{all_maps}
  \end{figure*}

In Figure \ref{shielding_lengths}, we present the distribution of shielding lengths measured in the simulation for the four different cases we test. The first is the smoothing length of the gas resolution elements, which is a commonly used approximation for $L_{\rm shield}$. A caveat of this definition is that the shielding length will be a function of the resolution of the simulation. In particular, in the limit of infinite resolution, the smoothing length will become very small and our model would approach the optically thin limit. We nevertheless include it in our analysis as it is a commonly used approach. In the FIRE-2 simulations, this length scale is also identical to the gravitational force softening of the gas and so sets the spatial resolution of the particle. It can be written as
\begin{equation}
  L_{\rm Sm} = 16 \, {\rm pc} \, \left( \frac{m_{\rm gas}}{1000 M_{\odot}}\right)^{1/3} \, \left(\frac{n_{\rm H}}{10 {\rm cm}^{-3}}\right)^{-1/3},
    \end{equation}
where $m_{\rm gas}$ is the mass resolution and $n_{\rm H}$ is the hydrogen number density. The second length scale we consider is the Jeans length, which is defined as
\begin{equation}
  L_{\rm Jeans} = \sqrt{\frac{\pi \, c_{\rm s}^2}{G \, \rho}}, 
  \end{equation}
where $L_{\rm Jeans}$ is the Jeans length, $c_{\rm s}$ is the sound speed, $G$ is the gravitational constant and $\rho$ is the density of each resolution element. This corresponds to the length scale at which a system becomes self-gravitating \citep[e.g.,][]{schaye2008}. The third and fourth cases we test are based on a Sobolev-like length scale defined using local density gradients \citep{sobolev1957,gnedin2009}. This definition of the Sobolev length is given by
\begin{equation}
\label{eqn:lsob}
  L_{\rm Sob} = \frac{\rho}{|\nabla \rho|}.
\end{equation}
We also explore the effect of using the length scale that is used in \textsc{gizmo} to convert densities to surface densities. These surface densities are used in the \citet{krumholz2011} fitting function to calculate H$_{2}$ fractions, as well as to attenuate the background radiation of local sources when determining the radiative feedback \citep{hopkins2018fire2}. This is a combination of the Sobolev length defined in equation \ref{eqn:lsob}, plus a contribution to the shielding length from the local cell. This is defined as
\begin{equation}
  L_{\rm Sob+cell} = \frac{\rho}{|\nabla \rho|} + \frac{L_{\rm sm}}{N_{\rm ngb}^{1/3}},
\end{equation}
where $N_{\rm ngb}$ is the number of nearest neighbours. This additional term is an approximation for the interparticle distance, which is smaller than the smoothing length of the cell. As for the local UV fluxes, we output this directly from the simulation when restarting the snapshot. We note that none of these length scales are equivalent to the velocity-gradient length scale we are using in \textsc{radmc-3d} to calculate the level populations, which make a moderate difference to the results \citep{penaloza2018}. As demonstrated in Figure \ref{shielding_lengths}, although the medians of the distributions of shielding lengths are similar, the detailed shapes can be quite different.

\begin{figure*}
  \includegraphics[width=2\columnwidth]{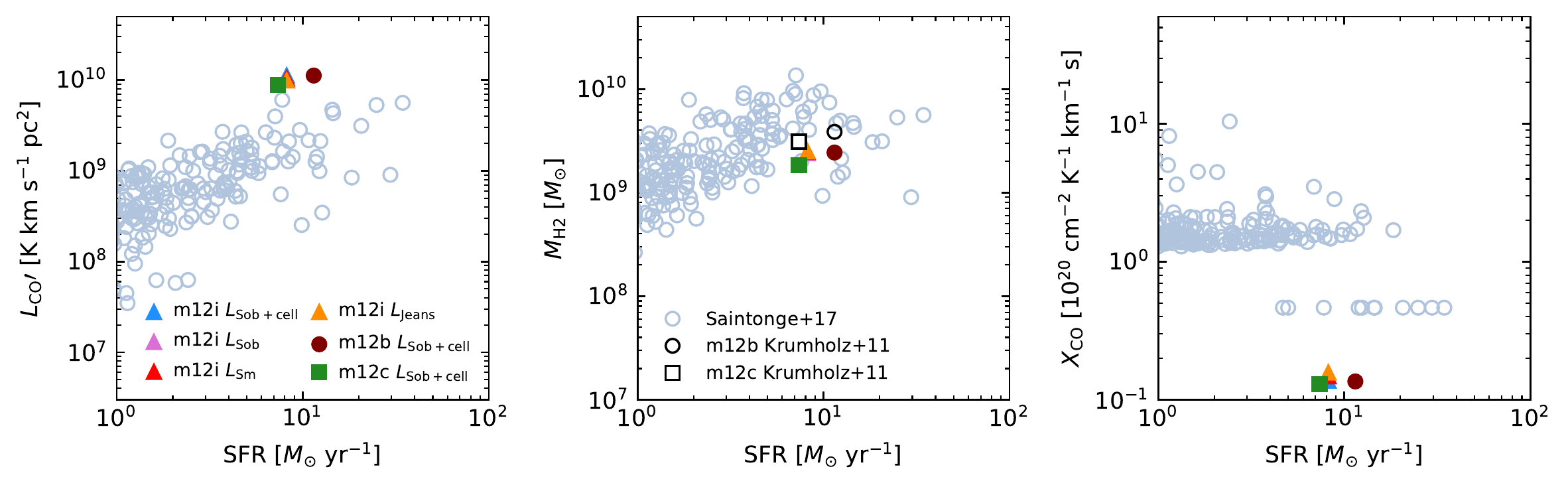}
  \caption{CO(1-0) luminosity (left), H$_{2}$ gas mass (middle) and CO-to-H$_{2}$ conversion factor (right) at a given star formation rate. Plotted in light blue are the observations from the xCOLD GASS sample \citep{saintonge2017} assuming the metallicity-dependent CO-to-H$_{2}$ conversion factor from \citet{accurso2017}. The blue, orange, red and pink triangles show the results for the m12i galaxy assuming different shielding lengths. Note that these points overlap with each other almost perfectly. The brown circle and green square show results from other simulations, assuming the Sobolev shielding length plus a contribution from the local cell. In the middle panel, the black open points are obtained from calculating the H$_{2}$ fraction for each particle using the \citet{krumholz2011} fitting function (which is used to determine the molecular fraction of gas in FIRE-2).}
  \label{lco_obs}
    \end{figure*}

We use these different assumptions about the shielding lengths as input for \textsc{chimes} and calculate the resulting CO abundances for each particle, and hence the total CO(1-0) luminosity for the galaxy.  Results for three of our shielding lengths are shown in Figure \ref{all_maps}. Looking first at the case where we use the Sobolev length plus a contribution from the local cell, we find that the results are unsurprising: the H$_{2}$ and CO clearly trace the spiral arms of the galaxy (first and second rows). The spiral arms are also home to the young stars, and hence where the UV flux is highest (Figure \ref{gal_maps}). This demonstrates the importance of accounting for the shielding from the UV background and local sources. In the third row of Figure \ref{all_maps}, we show the velocity-integrated CO(1-0) brightness temperature ($W_{10}$). This is calculated as
\begin{equation}
W = \frac{1}{2 k_{\rm B}} \left( \frac{c}{\nu} \right)^2 \int I_{\nu} \mathrm{d}v,
\end{equation}
where $k_{\rm B}$ is the Boltzmann constant, $c$ is the speed of light, $\nu$ is the frequency, $I_{\nu}$ is the line intensity computed by \textsc{radmc-3d} and $v$ is the velocity. It is clear that the emission is dominated by the high column density regions, and also that there is a significant fraction of H$_{2}$ that is ``CO-dark''. In the fourth row we show the CO-to-H$_{2}$ conversion factor ($X_{\rm CO}$), defined as in equation \ref{eqn:xco}. This is low throughout the disk compared to the Milky Way value (here we find an emission-weighted $X_{\rm CO} \sim 10^{19} \, {\rm cm}^{-2} \, {\rm K}^{-1} \, {\rm km}^{-1} \, {\rm s}$, a factor of ten smaller than expected). This discrepancy will be further quantified and discussed in more detail below. 

Next comparing the different columns, we investigate the effect of changing the definition of shielding length (columns 1--3) or increasing the cosmic ray ionization rate. We do not see a large difference between choosing the Sobolev length plus a contribution from the local cell, Jeans length or smoothing length (perhaps because the median values are similar; Figure \ref{shielding_lengths}). The most striking difference arises when we increase the cosmic ray ionization rate by a factor of 10. This dramatically decreases the CO abundance and hence the CO(1-0) emission, while only having a moderate effect on the  H$_{2}$ abundance. This increases the amount of ``CO-dark'' gas and therefore raises the CO-to-H$_{2}$ conversion factor. This will be discussed further in Section \ref{sec:params}.

\subsection{Comparison with observations}

We next wish to check how these models compare with observations of CO in nearby galaxies (Figure \ref{lco_obs}). In the first panel, we plot the CO(1-0) luminosity as a function of star formation rate and compare with the observed xCOLD GASS sample \citep{saintonge2017}. For all cases of our assumed shielding lengths, the CO(1-0) luminosity appears to be overestimated by almost an order of magnitude. We also find very little difference between the different definitions of shielding length, with the CO(1-0) luminosity varying by only five per cent depending on the definition that we use. Likewise, we see similar results for different zoom simulations of Milky Way-mass galaxies (the m12b and m12c simulations), highlighting that this issue is not specific to an individual simulation. We note that we are plotting the instantaneous star formation rate in the simulation here, which we compute by summing the star formation rate over all the gas in the disc, and that this may not be a fair comparison with the observations \citep[e.g.,][]{hayward2014,sparre2015}. However, this effect is unlikely to be significant enough to explain our overestimates of the CO(1-0) luminosity, and m12i in particular has a very stable recent star formation history, with an instantaneous star formation rate that is close to the average over the last 100 Myr.

In the middle panel, we plot the H$_{2}$ gas masses we find in the simulations (filled triangles). We also show the observational estimates, but emphasise that these are derived quantities and that a conversion factor must be assumed (shown here in the right panel). We find that our H$_{2}$ masses are somewhat low at a given star formation rate, but they do fall within the scatter of the observations. For comparison, we also show the H$_{2}$ masses estimated using the \citet{krumholz2011} method, which is used in the simulations to calculate the molecular fraction of gas. 

 We find that our H$_{2}$ masses are consistently smaller than the \citet{krumholz2011} estimate by about a factor 2, although these also fall towards the lower end of the H$_2$ masses inferred from observations. This is consistent with what was found in \citet{orr2018}, which showed that these simulations somewhat underestimate the surface density of cold and dense gas at a given star formation rate surface density.  Part of the difference between the \citet{krumholz2011} estimates and our chemical modelling could be due to differences in the assumed radiation field. When computing the \citet{krumholz2011} estimate, we assume the radiation field is a function of density and metallicity, which may differ from the inhomogeneous UV background computed in the FIRE-2 simulations that we are using in our chemical modelling (see the right panel of Figure \ref{gal_maps}). The mean amplitude of this inhomogeneous UV background is 1.04 $G_0$ in m12i. This is close to the Milky Way average, and perhaps somewhat lower than expected given that the star formation rate in the simulation is a few times higher than the Milky Way value. We find that the amplitude of this background is highly correlated with density (Figure \ref{all_maps}) as assumed in the \citet{krumholz2011} model.

\begin{figure*}
  \includegraphics[width=2\columnwidth]{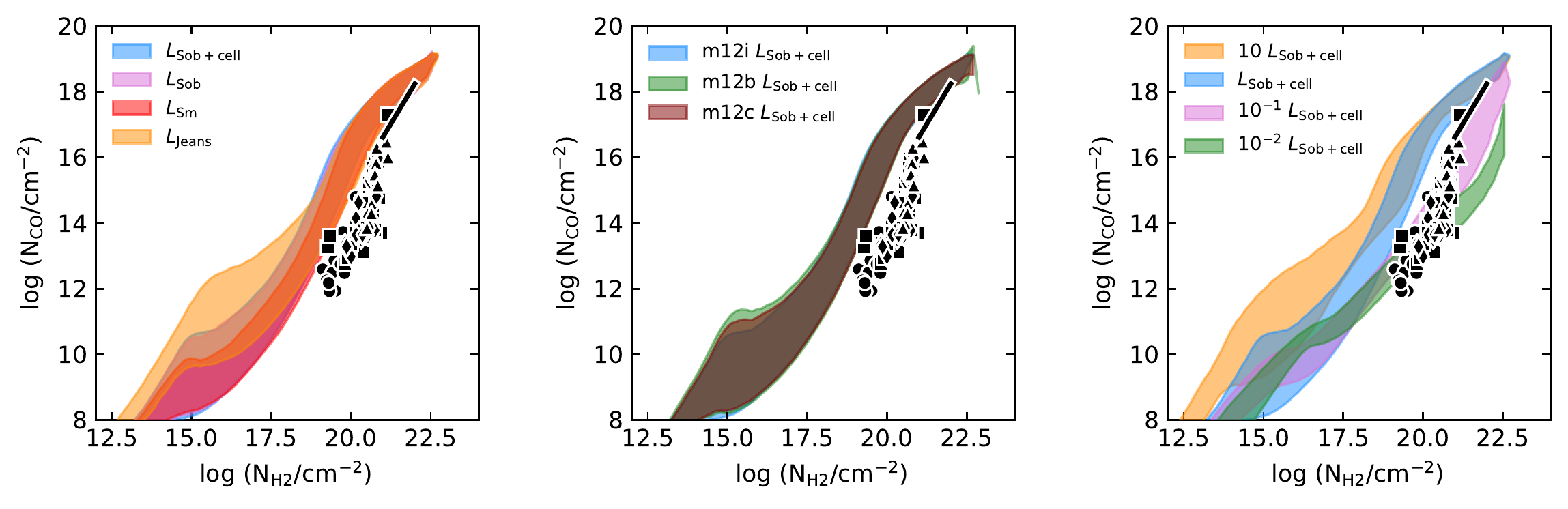}
\caption{Relation between CO and H$_{2}$ column densities for different definitions of shielding length (left), different haloes (middle) and models where we have multiplied the distribution of shielding lengths by a constant factor (right). The shaded regions show the range spanned by the 15$^{\rm th}$ and 85$^{\rm th}$ percentiles of the models. Also plotted are constraints from UV absorption lines in the Milky Way including a fit to a compilation of  observations from \citet{federman1990} (solid line), as well as observations from \citet{rachford2002} (triangles), \citet{crenny2004} (circles), \citet{sheffer2008} (squares) and \citet{burgh2010} (diamonds).}
  \label{nh2nco}
    \end{figure*}

In the right panel of Figure \ref{lco_obs}  we compare the CO-to-H$_{2}$ conversion factor assumed for the xCOLD GASS sample using the metallicity-dependent CO-to-H$_{2}$ conversion factor determined by \citet{accurso2017}.  The CO-to-H$_{2}$ conversion factor plotted for our models here is an emission-weighted average, defined following \citet{narayanan2012} as
\begin{equation}
\langle X_{\rm CO} \rangle = \frac{ \int \Sigma_{\rm H2} \, \mathrm{d}A}{ \int W_{10} \, \mathrm{d}A},
\end{equation}
where $\Sigma_{\rm H2}$ is the H$_2$ surface density. We find that our models have a CO-to-H$_{2}$ conversion factor that is too low by almost an order of magnitude compared to what is assumed in the \citet{accurso2017} estimates. It is even lower than the value for merging galaxies assumed in the xCOLD GASS sample ($4.7 \times 10^{19} \, {\rm cm}^{-2} \, {\rm K}^{-1} \, {\rm km}^{-1} \, {\rm s}$). While there are expected to be trends in the conversion factor based on environmental parameters \citep{sandstrom2013}, the conversion factor we recover here is more consistent with values usually assumed for ULIRGs \citep{bolatto2013} suggesting that something is missing in our chemical modelling. This will be investigated in more detail in section \ref{sec:shieldinglen}. Using the shielding length favoured by \citet{safranekshrader2017} (the Jeans length with a temperature cap of 40 K) did reduce the CO(1-0) emission, but only by a factor of two. Accounting for dust depletion would further reduce the CO(1-0) emission by another factor of two.  This would bring the CO(1-0) luminosity into better agreement with the observations (although still towards the high end), but would still underpredict the CO-to-H$_{2}$ conversion factor. 

As mentioned above and demonstrated in the middle panel of Figure \ref{lco_obs}, we found that our H$_{2}$ masses are reasonably consistent with the data. We further find that the radial profile of the H$_{2}$ surface density is in good agreement with the measurement of \citet{mivilledeschenes2017} in the central 10 kpc of the galaxy, and in fact overpredicts the  H$_{2}$ surface density by almost an order of magnitude at larger radii. This excess of H$_{2}$ surface density at large radii is also present when we assume the \citet{krumholz2011} relation for calculating the molecular fraction. We further find that the atomic to molecular transition in our simulations is in agreement with observations from \citet{gillmon2006}, \citet{wolfire2008} and \citet{rachford2009}. As increasing the H$_{2}$ abundance further would break the agreement with the observations of the H$_{2}$ surface density at small radii, this suggests that our models have a CO(1-0) luminosity that is too high for their H$_{2}$ gas mass.

\begin{figure*}
  \includegraphics[width=2\columnwidth]{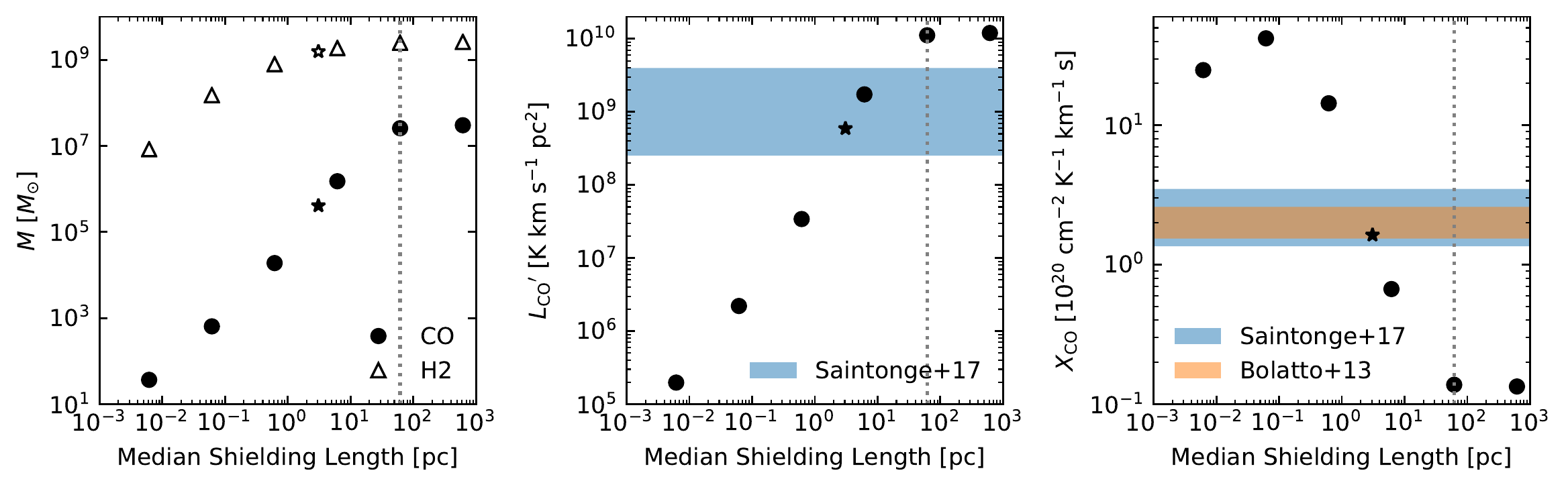}
  \caption{The effects of changing the shielding length on the CO and H$_{2}$ masses (left), the CO(1-0) luminosity (middle) and the emission-weighted CO-to-H$_{2}$ conversion factor (right). The shielding lengths here are calculated using the Sobolev length plus a contribution from the local cell, rescaled by a constant factor. The grey dotted vertical line shows median shielding length of the original distribution. The shielding length plotted here is the median in shielding length for cold ($T < 300$ K) gas in the disc. The star represents a model that falls within the observed scatter in CO(1-0) luminosity and CO-to-H$_{2}$ conversion factor  for the star formation rate in the simulation. This model is equivalent to 5 per cent of the distribution of Sobolev lengths plus a contribution from the local cells, which gives a median shielding length of 3 pc in the cold gas. We compare with the observations from the xCOLD GASS survey \citep{saintonge2017} and \citet{bolatto2013}.}
  \label{co_shielding}
    \end{figure*}

\subsection{Modelling the CO-to-H$_{2}$ conversion factor}

We next test the CO and H$_{2}$ abundances in our models against absorption line measurements at UV wavelengths of the CO and H$_{2}$ column densities. This allows us to neglect the radiative transfer effects, and isolate any effects in our chemical modelling. To measure the column densities in the simulations, we map the CO and H$_{2}$ masses of each particle onto a 2048$^{2}$ regular cartesian grid (chosen to be at the same spatial resolution as the line emission maps we produce with \textsc{radmc-3d}). We interpolate assuming a cubic spline, rotate the galaxy to be face-on, and treat each pixel of this map as a single sightline from which we take the CO and H$_{2}$ column densities. The results of this are shown in Figure \ref{nh2nco}. In the left panel, we show the CO and H$_{2}$ column densities we measure in the Jeans length, smoothing length, Sobolev length and the length scale used in the FIRE-2 simulations ($L_{\rm Sob+cell}$) compared to the absorption line data. We find that, consistent with Figure \ref{lco_obs}, we seem to overproduce CO at a given H$_{2}$ abundance. In particular, the point where the CO column density begins to steeply rise appears to begin at a lower H$_{2}$ column density in our models than what is favoured by the observations. This leads to CO column densities that are too high for their equivalent H$_{2}$ column density compared with the observations.  The same is true for different haloes in the FIRE-2 suite of simulations (shown for the Sobolev length plus local cell correction model, $L_{\rm Sob+cell}$, in the middle panel), confirming that this is an issue in the chemical modelling and is not specific to the properties of the m12i simulation.

After exploring the parameter space of assumptions in our modelling (discussed in more detail below), we found that by far the most effective way to reduce the CO abundance while not strongly changing the H$_{2}$ abundance was by assuming a smaller shielding length. This is perhaps indicative of unresolved substructure in the gas distribution. We parameterised this in terms of a constant factor times the density-based Sobolev length plus the cell correction ($L_{\rm Sob+cell}$) measured for each resolution element, the same length scale used to convert volume densities to column densities in the hydrodynamic simulation. Reducing the shielding length means that at a given volume density, the associated column density will be smaller. Since CO primarily depends on being shielded from photodissociation to form, while H$_{2}$ is more sensitive to the volume density and metallicity \citep[e.g.,][]{glover2011}, changing this shielding length changes the ratio of CO to H$_{2}$. 

As shown in the right panel of Figure \ref{nh2nco}, lowering the shielding length has the desired result of reducing the abundance of CO in our models at a given H$_{2}$ column density. As the shielding length is decreased, the point at which CO can form effectively is shifted to increasingly higher H$_{2}$ column densities. Decreasing the shielding length further then leads us to undershoot the CO column densities at large H$_{2}$ column densities. It is therefore possible to scale the distribution of shielding lengths in our simulation such that we can find good agreement with the absorption line data.  This smaller shielding length can perhaps be thought of as a subgrid model for unresolved structure in the density field of GMCs in the simulation, as discussed in section \ref{interpretation}.

\begin{figure*}
  \includegraphics[width=2\columnwidth]{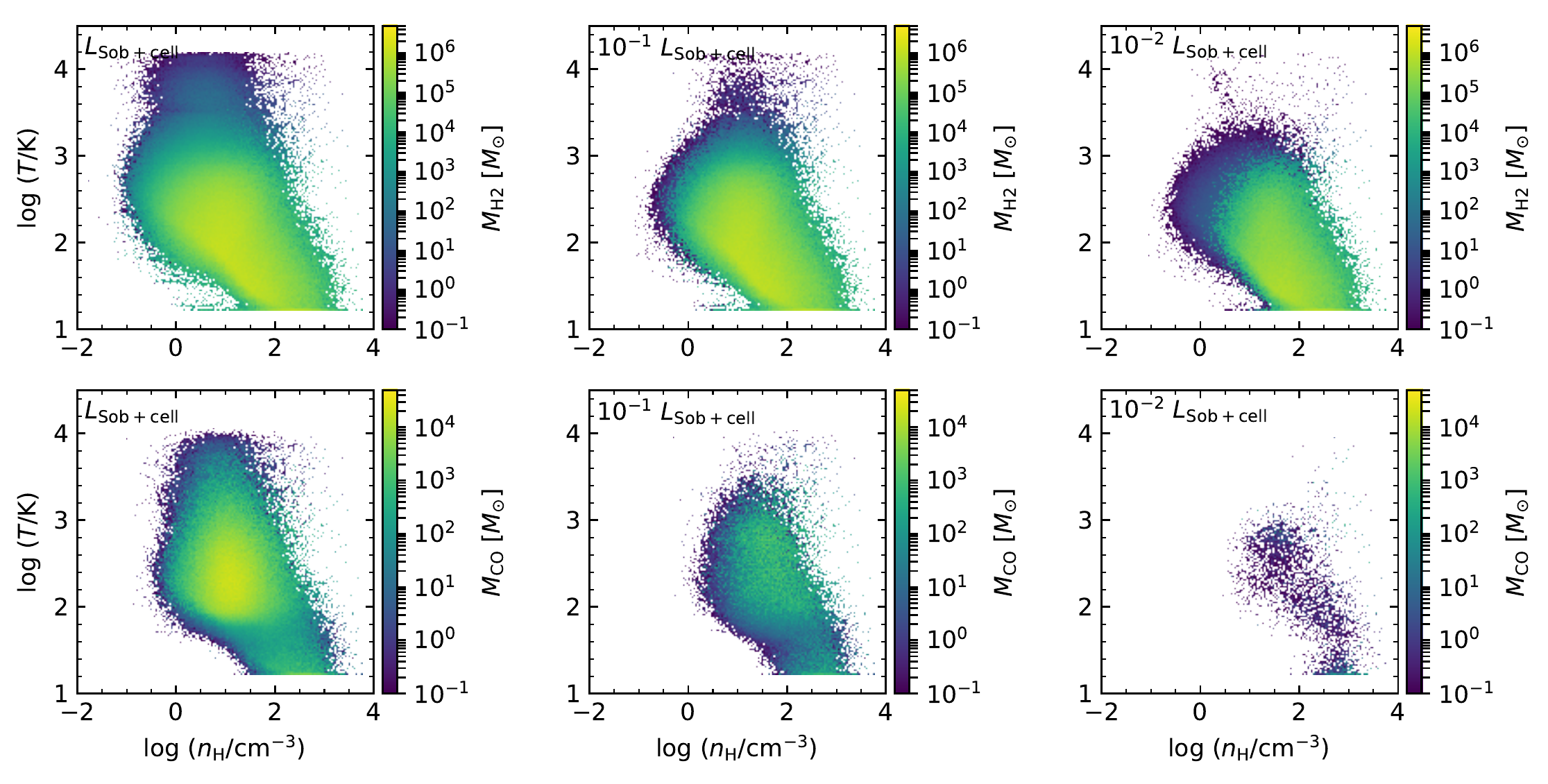}
  \caption{We show the temperature and hydrogen number density of all gas particles that contain at least 0.1 $M_{\odot}$ of H$_{2}$ (top panel) and CO (bottom panel). When computing the temperature we assume a mean molecular weight $\mu=2$ appropriate for molecular gas. Each column shows the distribution of shielding lengths rescaled by a different factor. The colour of the points reflects the mass of that species in a given bin. From left to right: assuming that the shielding length is the Sobolev length plus a correction from the local cell ($L_{\rm Sob+cell}$), 0.1 times $L_{\rm Sob+cell}$ and 0.01 times $L_{\rm Sob+cell}$. These length scales correspond to median shielding lengths 61.6 pc, 6.16 pc and 0.616 pc in cold ($T < 300$ K) gas. As the shielding length is reduced, we see that the mass of CO strongly declines. The H$_{2}$ mass also declines, but not as steeply.}
  \label{co_h2_temp_dens}
    \end{figure*}

\subsection{Effect of decreasing the shielding length}
\label{sec:shieldinglen}

In this section, we explore the effects of changing the shielding length on the CO and H$_{2}$ masses, the CO(1-0) luminosity and the CO-to-H$_{2}$ conversion factor. As above, we change the distribution of shielding lengths by rescaling the distribution of $L_{\rm Sob+cell}$ (the same shielding lengths used in the simulation) by a constant value, i.e. each resolution element has its shielding length reduced by the same factor. This was the simplest way to implement this, but in reality the factor we multiply by may depend on the local properties of the gas. The results of changing this shielding length are shown in Figure \ref{co_shielding}. In the left panel of Figure \ref{co_shielding}, we show how the total CO and H$_{2}$ masses in the simulation are changed by making the shielding length smaller. Above a certain threshold, there is no effect as all the gas that has a high enough density is fully self-shielded. As the shielding length is decreased further, both the CO and H$_{2}$ masses begin to decline. However, the CO is more sensitive to this effect and the CO mass declines faster than the H$_{2}$ mass. This allows us to reduce the CO abundance in our model without changing the H$_{2}$ abundance by much, exactly as required.

\begin{figure*}
  \includegraphics[width=2\columnwidth]{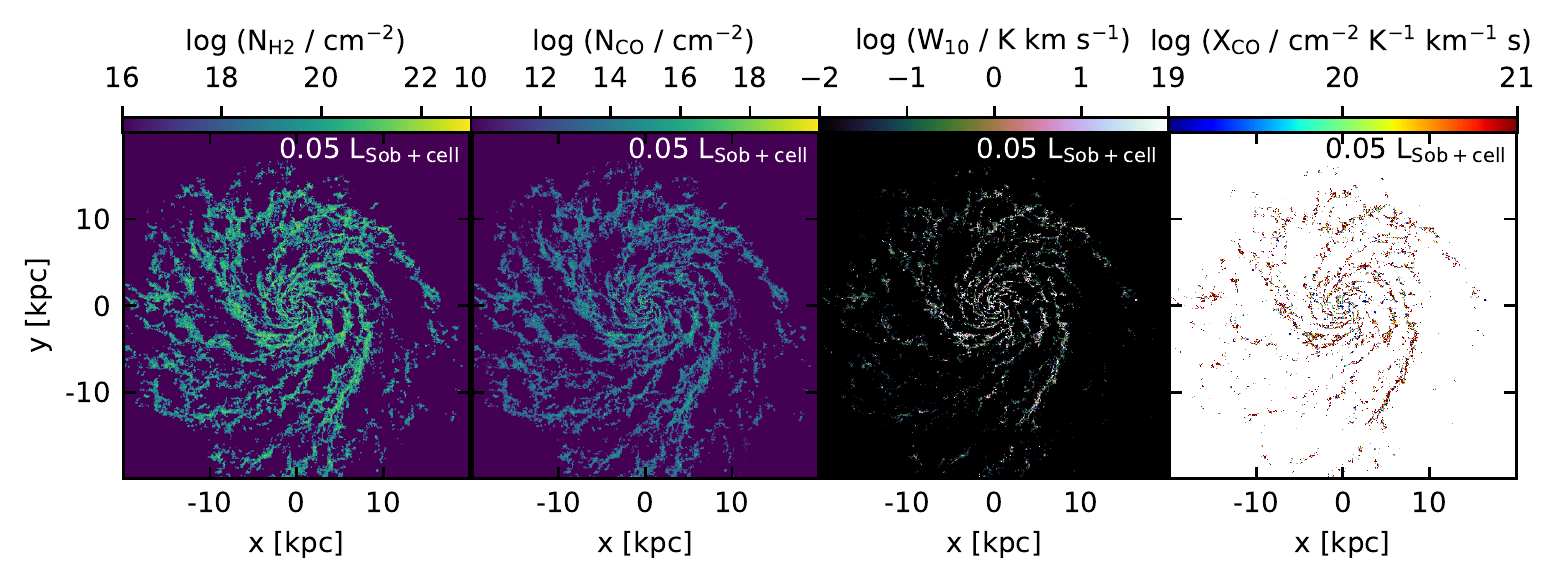}
  \caption{From left to right: For a shielding length of 0.05 $L_{\rm Sob+cell}$, a model that has integrated CO(1-0) luminosity and emission-weighted CO-to-H$_{2}$ conversion factor in line with observations, given its star formation rate, we show maps of the H$_{2}$  column density, the CO column density, the velocity-integrated CO(1-0) brightness temperature and the CO-to-H$_{2}$ conversion factor for pixels that have $W_{10} > 0.1$ K km s$^{-1}$ in the m12i simulation.}
  \label{maps}
  \end{figure*}

\begin{figure*}
  \includegraphics[width=2\columnwidth]{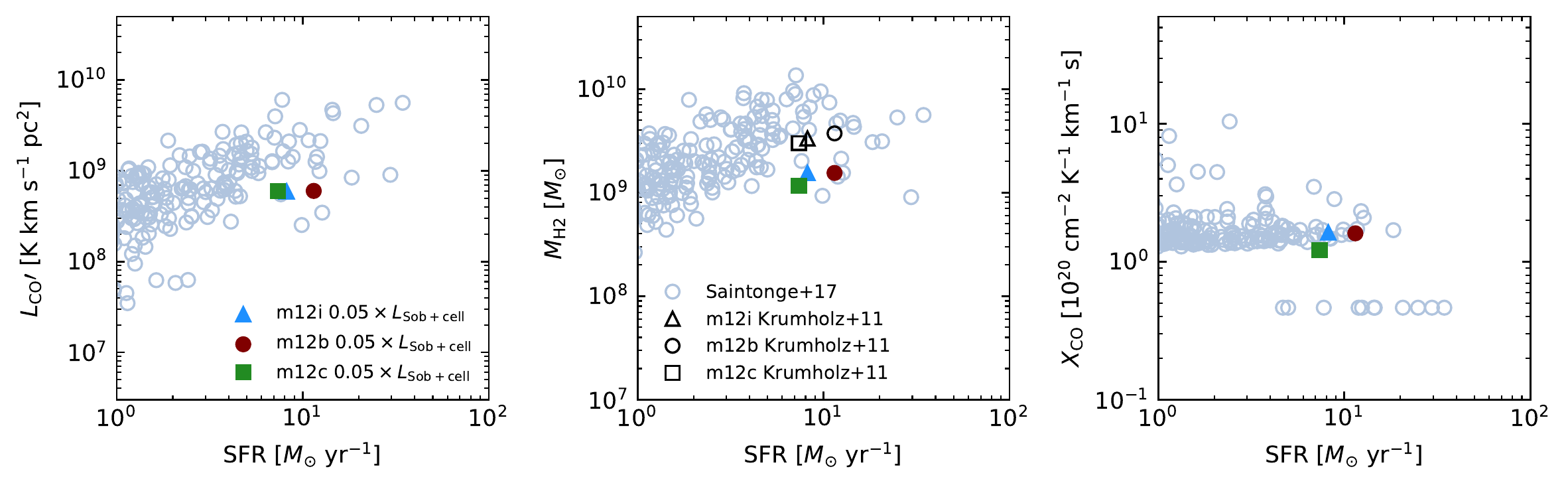}
  \caption{CO(1-0) luminosity (left), H$_{2}$ gas mass (middle) and CO-to-H$_{2}$ conversion factor (right) at a given star formation rate. Plotted in light blue are the observations from the xCOLD GASS sample \citep{saintonge2017} assuming the metallicity-dependent CO-to-H$_{2}$ conversion factor from \citet{accurso2017}. The coloured points are results from the simulations all assuming that the distribution of Sobolev shielding lengths plus local cell correction has been rescaled by a constant factor of 0.05, corresponding to a median length of 3 pc for cold ($T < 300$ K) gas. In the middle panel, the black open points are obtained from calculating the H$_{2}$ fraction for each particle using the \citet{krumholz2011} fitting function.}
  \label{lco_obs_len}
    \end{figure*}

\begin{figure*}
  \includegraphics[width=2\columnwidth]{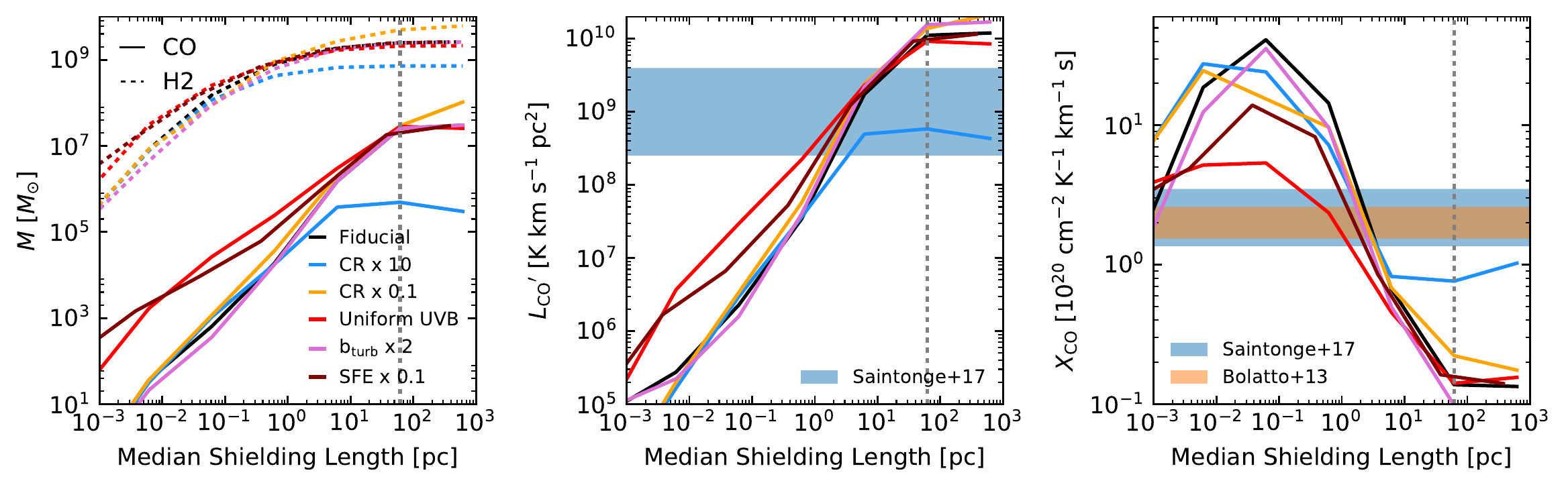}
  \caption{The effects of the changing the shielding length on the CO and H$_{2}$ masses (left), the CO(1-0) luminosity (middle) and the emission-weighted CO-to-H$_{2}$ conversion factor (right). The grey dotted vertical line shows the median of the original distribution of Sobolev lengths plus local cell correction. The black lines are the same as the black points in Figure \ref{co_shielding}. The other coloured lines show the effects of changing different assumptions in our modelling: increasing the cosmic ray ionization rate by a factor 10 (blue line), lowering the cosmic ray ionization rate by a factor 10 (orange line), using a homogeneous UV background set to the interstellar radiation field from \citet{black1987} (red line), increasing the subgrid turbulence in our model by a factor 2 (pink line) and restarting the simulation with a lower local star formation efficiency to allow more dense gas to build up (maroon line). We compare with the observations from the xCOLD GASS survey \citep{saintonge2017} and \citet{bolatto2013}.}
  \label{co_params}
    \end{figure*}

A similar effect is seen in the CO(1-0) luminosity (middle panel). For the unscaled Sobolev lengths with local cell correction, we recover our result from above that we overpredict the CO(1-0) emission compared to the observations of \citet{saintonge2017}. As the shielding length is decreased, the CO(1-0) luminosity falls as less CO is now being formed. For a small enough shielding length, we underproduce the CO(1-0) emission compared to the observations. The behaviour of the CO-to-H$_{2}$ conversion factor is slightly more complicated. For large shielding factors, we underpredict the CO-to-H$_{2}$ conversion factor favoured for the Milky Way by \citet{bolatto2013}. As the shielding length is decreased, the CO-to-H$_{2}$ conversion factor rises, as we are producing less CO in the models while the H$_{2}$ abundance stays relatively constant. Eventually, the relation between shielding length and CO-to-H$_{2}$ conversion factor turns over, as for small enough shielding lengths the production of H$_{2}$ is also affected. However, this occurs in the regime where the CO(1-0) emission is underpredicted, and is therefore not important here. 

To build some intuition for what is happening when we decrease the shielding length, we plot the temperature and density covered by the H$_{2}$ and CO in for three different cases of the rescaled shielding length distribution  (Figure \ref{co_h2_temp_dens}). This shows that the gas phase traced by CO changes significantly with the assumed shielding length. For the model that uses the unscaled Sobolev shielding length plus local cell correction, we find that CO exists in gas with densities as low as $n_{\rm H}=1$ cm$^{-3}$ and temperatures as high as $T=10^{4}$ K. As the shielding length decreases, the CO in particular is restricted to increasingly higher densities and lower temperatures. This is also true for the H$_{2}$, but the effect is not as strong.

By rescaling the shielding length by an appropriate factor, it is therefore possible to ``tune'' our model to match both the observed CO(1-0) luminosity and estimated CO-to-H$_{2}$ conversion factor at fixed star formation rate. One such model is shown in Figure \ref{co_shielding} and is represented by the star. This model rescales the distribution of Sobolev lengths with the local cell correction by a factor of 0.05, which results in a median shielding length of 3 pc in the cold ($T < 300$ K) gas in the disc. It is reassuring that it is possible to find a rescaled shielding length distribution that simultaneously reproduces the expected CO(1-0) luminosity and estimated CO-to-H$_{2}$ conversion factor. Maps of the resulting H$_{2}$ and CO column densities, CO(1-0) brightness temperature and CO-to-H$_{2}$ conversion factor for the model with 0.05 $L_{\rm Sob+cell}$ are shown in Figure \ref{maps}. In contrast to the panels of Figure \ref{all_maps} that use the same cosmic ray ionization rate, we now find a lower CO column density and much of the diffuse CO outside the spiral arms has been removed.  Likewise the CO(1-0) emission is now restricted to the spiral arms of the galaxy. We also find a higher CO-to-H$_{2}$ conversion factor throughout the disc, with the regions that are brightest in CO(1-0) emission showing a conversion factor that is in line with observational estimates \citep[e.g.][]{bolatto2013}. Using this same value for the reduced shielding length also produces a similar level of agreement between the data and other simulations (Figure \ref{lco_obs_len}).

\subsection{Effects of changing other parameters}
\label{sec:params}

There are of course other assumptions that we make when calculating the chemical abundances, and we investigate this further in Figure \ref{co_params}. As in Figure \ref{co_shielding}, we show how the  CO and H$_{2}$ masses, the CO(1-0) luminosity and the CO-to-H$_{2}$ conversion factor change as the shielding length is varied. We present models that make different assumptions compared to our fiducial model (black line). First, we explore the effects of changing the cosmic ray ionization rate, raising/lowering it by a factor 10 (blue and orange lines). Some observations suggest that the cosmic ray ionization rate could be a factor 10 lower \citep{williams1998}. We found that using a lower cosmic ray ionization rate increased the CO and  H$_{2}$ abundances by a small amount, and did not have a large effect on the CO(1-0) luminosity or CO-to-H$_{2}$ conversion factor.

Somewhat more successful was the model where we increased the cosmic ray ionization rate (see also Figure \ref{all_maps}). There is little evidence that the average value for the Milky Way should be higher, however it may scale with the star formation rate of the galaxy \citep{lacki2010}. The star formation rate in the simulation is almost a factor four higher than estimates for the Milky Way \citep[e.g.][]{chomiuk2011}. which provides some motivation that the cosmic ray ionization rate could be higher. Apart from changing the shielding length, we find that  increasing the cosmic ray ionization rate  was the most effective way to produce less CO in the models. This is because cosmic rays are able to penetrate the shielded regions and destroy the CO via a reaction with \ion{He}{II} \citep{bisbas2015,narayanan2017}. Increasing the cosmic ray ionization rate has however a smaller effect on the H$_{2}$, although the H$_{2}$ abundance is still decreased. We found that, when using the unscaled Sobolev length plus local cell correction as our shielding length, increasing the cosmic ray ionization rate resulted in a CO(1-0) luminosity that was compatible with the observations. However, we do not favour this solution as the emission-weighted CO-to-H$_{2}$ conversion factor we measure was still too small by a factor of a few. 

Next, we explore the effects of using a homogeneous UV background set to the interstellar radiation field from \citet{black1987}. This is obviously not a realistic case, compared with our fiducial model for the UV background which accounts for the effects of local sources, and in reality we expect at least some of the GMCs to be sitting near young stars and therefore in regions of enhanced UV flux (see, e.g., Figure \ref{gal_maps}). In practice however, we find little difference between our fiducial model (which includes a inhomogeneous UV background) and this model with a uniform UV background. This is because the choice of UV background makes little difference at high shielding lengths, as the CO and H$_{2}$ are so effectively shielded that their abundances are almost independent of the assumed UV flux. At lower shielding lengths, we see that the CO abundances are increased when the uniform background is used compared with the inhomogeneous case due to the overall lower amplitude of the UV background. However, these differences only arise when the CO(1-0) luminosities are already underproduced compared to the observations.

We also explore the effect of changing the subgrid turbulence, which accounts for unresolved turbulent motion in the simulation. As described in section \ref{chimes}, our fiducial subgrid turbulence has a line width of 7.1 km s$^{-1}$, corresponding to a velocity dispersion of 5 km s$^{-1}$. This value is also used in \citet{krumholz2012} to model the \ion{H}{i} to H$_{2}$ transition and its implications for star formation, and is close to the observed velocity dispersion in nearby GMCs. We test the effect of increasing this Doppler parameter by a factor 2 (pink line, Figure \ref{co_params}), but find it only has a minimal effect.

The final parameter we vary is the local star formation efficiency in the underlying cosmological simulation. This parameter sets the fraction of molecular gas that turns into stars per free fall time. The default FIRE-2 physics assumes that this efficiency is 100 per cent in dense star-forming gas, as the feedback automatically regulates the star formation without having to force a lower efficiency by hand. \citet{hopkins2018fire2} show that assuming that the local star formation efficiency in dense gas is 100 times lower or higher has a negligible effect on the star formation rate history of the galaxy. Lowering the star formation efficiency in dense star-forming gas however does have an effect on the amount of dense gas present in the simulation: if it is lowered, then more dense gas is accumulated to achieve the same star formation rate. As we are particularly interested in the dense, molecular phase it is worth investigating this parameter. We restarted a snapshot using a star formation efficiency in dense star-forming gas $\epsilon =0.1$ (compared to the fiducial $\epsilon =1$) and reran the simulation for 900 Myr down to redshift zero to allow the dense gas to build up. Looking at the density PDF, we did see a increased amount of dense gas in the distribution compared to the fiducial run. However, changing this efficiency parameter only affected a small fraction of resolution elements in the simulation. This is reflected in our CO and H$_{2}$ models which are very similar to the fiducial case at high shielding lengths, although an increase in  CO and H$_{2}$  abundances is seen at lower shielding lengths. We therefore find that our results are not sensitive to the local star formation efficiency we assume, as changing this parameter only impacts a small fraction of the gas resolution elements.

In summary, although changing the default parameters in our modelling can affect the CO and H$_{2}$ abundances we predict, these effects are generally small at fixed shielding length (Figure \ref{co_params}). Apart from the shielding length, we found that the most significant parameter was the cosmic ray ionization rate. However, at the fiducial Sobolev plus local cell shielding length, increasing the cosmic ray ionization rate did not produce a CO-to-H$_{2}$ conversion factor that agrees with the expected Milky Way value. We therefore conclude that changing the shielding length is the most effective mechanism we found for producing CO and H$_{2}$ abundances that are in agreement with the observations, although we note that there could also be alternative methods that may produce the same results.

\section{Discussion}
\label{interpretation}

By tuning the shielding length to reduce the CO(1-0) luminosity in the simulation, we have introduced a model that matches the observations quite well.  This model appears to recover the expected global molecular gas properties of the galaxy, such as its integrated luminosity and emission-weighted CO-to-H$_2$ conversion factor. Reducing the shielding length may not be the only way to recover these properties, and there may be a combination of different parameters that we could vary to achieve the same result. However, varying the shielding length was the most effective solution that we found. There are two connected reasons that this is an effective strategy. In the models that use common approximations for the shielding length, there is an over-abundance of CO for a given amount of H$_2$ (see, e.g., Figure \ref{nh2nco}). This results in an emission-weighted CO-to-H$_2$ conversion factor that is too low, as most of the emission is coming from low density regions, and the CO-to-H$_2$ conversion factor is low at these densities due to the overproduction of CO. As we reduce the shielding length, we lower the abundance of CO at lower densities. This raises the CO-to-H$_2$ conversion factor in that region, but the densities responsible for the bulk of the CO emission also change. This changes the weighting in our quoted CO-to-H$_2$ conversion factors, and restricts the gas it is probing to denser regions where the CO-to-H$_2$ conversion factor was initially higher.

Accounting for the effects of dust depletion will likely reduce the CO(1-0) emission and allow us to adopt a somewhat larger shielding length. However, we estimate that this will change things by a factor of two and we emphasise that matching the CO-to-H$_2$ conversion factor requires us to reduce the CO(1-0) emission by a factor of $\sim20$. We do note however that the effects of dust depletion are dependent on both the metallicity, density and molecular fraction of the gas \citep[see, e.g.,][]{chiang2018} and we intend to investigate this more carefully in future work. 

Decreasing the shielding length is effectively a ``subgrid'' model that we can tune to compensate for the finite resolution of the simulation. It is therefore interesting to think about the substructure that is not resolved in these simulations, and how it may relate to this model.  CO emission comes from GMCs, which are known to have extensive substructure that may effect the optical depth in that region. In fact, there is some observational evidence that optical depths in star forming regions may be lower than expected. The free-free emission in our galaxy is dominated by the contribution from the extended low density region \citep[ELD;][]{mezger1978,guesten1982}. Using observations from \textit{WMAP}, \citet{murray2010} showed that the ELD was associated with ionizing photons leaking out of massive star clusters rather than with photons coming from \ion{H}{ii} regions as previously thought \citep{lockman1976,anantharamaiah1985a,anantharamaiah1985b}. If ionizing photons are able to escape from star forming regions, then they must also be able to enter them, equivalent to lowering the optical depth in that region. By using a smaller shielding length scale in our model, we reduce the self-shielding experienced by each particle and effectively allow more of the radiation field to leak into each resolution element, lowering its optical depth.

An alternative interpretation is that the small shielding length we require corresponds to the ``coherence length'' of the gas. This length scale corresponds to the length over which the velocity of the gas changes by about the width of the shielding line. \citet{gnedin2011} and \citet{feldmann2012} have used a coherence length of 1 pc in their modelling of molecular gas, which allows them to successfully reproduce a range of observations, such as the molecular and atomic gas fraction as a function of gas column density and the dependence of the CO-to-H$_2$ conversion factor on metallicity. This value of 1 pc is close to the median of the distribution of rescaled shielding lengths in our preferred model. 

It is useful to place our work in context by comparing it to other studies of modelling CO in the literature. There are of course other published subgrid models for computing the CO emission from cosmological simulations \citep[e.g.,][]{krumholz2014des,olsen2016,narayanan2017,vallini2018}. In some respects, these models are more sophisticated than the simple model we have constructed here, and include subgrid models for the internal structure of gas resolution elements in the simulation based on properties such as the surface density of a cell or particle. The benefit of these subgrid models is that higher densities can be attained in the imposed substructure, which allows for the prediction of, e.g., higher order CO lines that would not be possible with the model we present here. The benefit of our modelling, however, is that it places a target length scale on the resolution that should be achieved in a cosmological simulation to self-consistently model CO(1-0) emission. Our work is therefore complementary to these other subgrid models, and we do not necessarily suggest it as the favoured approach.

We find that our preferred value of the median shielding length is also approaching the spatial scale of simulations where CO abundances can be computed without the aid of explicit subgrid modelling. Simulations of isolated galaxies, or segments of galaxies, run with chemical networks computed on-the-fly such as \citet{smith2014}, \citet{richings2016} and \citet{gong2018}, or run with chemical post-processing such as \citet{fujimoto2019} and  have successfully reproduced the expected value of the CO-to-H$_2$ conversion factor or have matched the observed CO-based Kennicutt-Schmidt relation (which would require a correct $X_{\rm CO}$). Even the lowest resolution of these simulations has a fixed spatial scale of 8 pc. This is almost a factor of eight times higher in resolution than the median smoothing length in the cold ($T$ < 300 K) gas in the simulations we study here (Figure \ref{shielding_lengths}), although the resolution in the densest regions in our simulations can be much higher. The fact that our preferred median shielding length of $\sim 3$ pc falls in the range of spatial resolution scales in the above works is perhaps indicative that our modelling is placing a limit on the resolution that must be attained to self-consistently model CO in cosmological simulations. However, we note that \citet{joshi2019} showed that scales of 0.04 pc are required for converged CO formation, which is still far below our preferred median shielding length scale.

Since our claim is that we need this small shielding length to compensate for the finite resolution of the simulations, we could test this with a higher resolution simulation that has a median spatial resolution of order  1 pc in the disk. In this case, our work predicts that the simulation should reproduce the observed CO(1-0) emission and CO-to-H$_{2}$ conversion factor using the smoothing length as the shielding length. This is a very ambitious resolution requirement, but the idea could perhaps be tested by restarting snapshots of existing simulations and splitting the particles. In this case the simulation only has to be run for a dynamical time to allow the gas to settle down, which is computationally more feasible. Such a run has already been performed for m12i, with the spatial resolution increased by a factor of two (mass resolution higher by a factor of eight). In our preliminary investigations we have found that the H$_2$ mass was increased by 1\%, but the CO mass decreased by 53\% compared to the original simulation. In both cases here we assumed the shielding length was equal to the smoothing length. This decrease is in line with what we would expect from our reduced shielding lengths, suggesting that it is indeed the resolution of the simulation that requires us to use a subgrid model. Another test would be to see how much ionizing radiation escapes from young star clusters in the hydrodynamic simulations. If little ionizing radiation escapes, in contrast to what is observed, this would suggest that the radiation is being over-attenuated due to the large shielding lengths assumed in the simulation. 

Another avenue of investigation is how well the simulations reproduce other emission lines. One possibility would be to look at higher order CO lines. However, these lines have increasingly high critical densities. The critical density of the $J=1-0$ transition is already at $\sim 10^{3}$ cm$^{-3}$, which is close to the highest density gas present in our simulation (although this critical density decreases with increasing optical depth). The $J=2-1$ critical density is almost an order of magnitude higher so our predictions for this line are unlikely to be accurate. Another possibility is to model line emission from other species, such as the [\ion{C}{ii}] 158 $\mu$m line. Our preliminary results from modelling the [\ion{C}{ii}] emission are that, as with the CO(1-0) emission, we overpredict the [\ion{C}{ii}] emission in a model that assumes the shielding length is equal to the Sobolev length plus the correction for the local cell. However, unlike the CO, we do not find better agreement with observations when we rescale the shielding length. This suggests that there may be still something missing in our modelling, and perhaps motivates a more explicit implementation of the ISM substructure \citep[e.g.][]{olsen2016,vallini2018,li2018}. We note that the [\ion{C}{ii}] 158 $\mu$m emission and CO(1-0) emission are probing very different phases, with the CO coming from GMCs while the [\ion{C}{ii}] can trace both ionized and neutral gas. We would therefore  not necessarily expect the same approach to reproduce both lines, and leave a more careful study to future work, preferably in simulations with higher mass resolution.

\section{Summary and Conclusions}

In this paper we have presented models of CO(1-0) emission in cosmological zoom simulations of Milky Way-mass galaxies which reproduce many of the observed properties of our Galaxy. We have used a chemical equilibrium solver to predict the H$_{2}$ and CO abundances, and a line radiative transfer code to find the emergent line intensity. We compare these models against observations, specifically testing them against CO(1-0) emission from galaxies in the xCOLD GASS survey \citep{saintonge2017} and UV absorption line constraints on CO and H$_{2}$ column densities along different lines of sight through the Galaxy. 

We explore the effect of changing many of the assumptions that go into our modelling, and find that our results depend most strongly on the shielding length we assume. This shielding length relates the volume densities in the simulation to the column densities required to calculate an optical depth for the gas. We investigate four definitions for the shielding length derived from  the simulation: the Jeans length, the smoothing length, a density-gradient based Sobolev length and the Sobolev length plus a correction for the local cell (which is used in the FIRE-2 simulations to estimate column densities). These definitions all produce a CO(1-0) luminosity that is too high given the star formation rate in the simulation. By reducing the shielding length, until the median of the distribution is 3 pc for the cold gas, we find that we can reproduce both the global CO(1-0) luminosity and CO-to-H$_{2}$ conversion factor expected for the simulated galaxies based on their star formation rates. However, we find that the  [\ion{C}{ii}] 158 $\mu$m emission is still overpredicted, suggesting that our modelling can be improved further in the future. Further progress will be therefore be made using higher resolution simulations, simultaneously modelling multiple emission lines and accounting for the effects of dust depletion. Comparing with spatial variations of these quantities within a galaxy may also provide new insights, e.g. comparing with the variations of the CO-to-H$_{2}$ conversion factor for individual GMCs as in \citet{narayanan2013}.

Our results suggest that it is important to account for the substructure in GMCs when making predictions for the molecular component of galaxies. This substructure occurs at a spatial scale as yet unresolved in the simulations we post-process here. The exquisite quality of existing and upcoming observations demand realistic theoretical models to aid the interpretation of the detailed ISM physics, and to relate the observable line emission to fundamental quantities such as the total molecular gas mass and star formation rate. However, it is interesting to explore simple models such as the one presented here, as they may provide some insight into the relevant physical processes.

\section*{Acknowledgements}

We thank Desika Narayanan and Xiangcheng Ma for their comments on this draft, and Yuxuan Yuan for his assistance in post processing the particle split snapshot. LCK acknowledges the support of a CITA fellowship and a Beatrice and Vincent Tremaine Fellowship. AJR was supported by a COFUND/Durham Junior Research Fellowship under EU grant 609412; and by the Science and Technology Facilities Council [ST/P000541/1]. We acknowledge the support of the Natural Sciences and Engineering Research Council of Canada (NSERC). This research was undertaken, in part, because of funding from the Canada Research Chairs program. CAFG was supported by NSF through grants AST-1517491, AST-1715216, and CAREER award AST-1652522; by NASA through grant 17-ATP17-0067; and by a Cottrell Scholar Award from the Research Corporation for Science Advancement. Support for PFH was provided by an Alfred P. Sloan Research Fellowship, NSF Collaborative Research Grant 1715847 and CAREER grant 1455342, and NASA grants NNX15AT06G, JPL 1589742, 17-ATP17-0214. AW received support from NASA, through ATP grant 80NSSC18K1097 and HST grants GO-14734 and AR-15057 from STScI, the Heising-Simons Foundation, and a Hellman Fellowship. DK was supported by NSF grant AST-1715101 and the Cottrell Scholar Award from the Research Corporation for Science Advancement. RF acknowledges financial support from the Swiss National Science Foundation (grant no 157591). Support for SRL was provided by NASA through Hubble Fellowship grant HST-JF2-51395.001-A awarded by the Space Telescope Science Institute, which is operated by the Association of Universities for Research in Astronomy, Inc., for NASA, under contract NAS5-26555. Computations were performed on the Niagara supercomputer at the SciNet HPC Consortium. SciNet is funded by: the Canada Foundation for Innovation; the Government of Ontario; Ontario Research Fund - Research Excellence; and the University of Toronto \citep{scinet_ponce,scinet_loken}.

\section*{Data Availability}

The data underlying this article will be shared on reasonable request to the corresponding author.

\bibliographystyle{mnras} \bibliography{xco_ref}

\bsp

\label{lastpage}

\end{document}